\documentclass[referee]{raa}            % referee version: for submission

\usepackage{graphicx,times}             %for PS/EPS graphics inclusion, new
\usepackage{natbib}
\usepackage{amssymb,amsmath}
\bibpunct{(}{)}{;}{a}{}{,}

\usepackage[pagebackref=true]{hyperref}

\begin{document}

  \title {High-resolution Observations of Clustered Dynamic Extreme-Ultraviolet Bright Tadpoles near the Footpoints of Corona Loops}
%   \subtitle{I. Place Your Subtitle Here}

\volnopage{Vol.0 (20xx) No.0, 000--000}      %%preserved for Editor. DOn't remove!
   \setcounter{page}{1}          %%starting page, preserved for Editor. DOn't remove!

   \author{Rui Wang %% Put your Chinese name in "( )" if you like. Note to open line 11 "\usepackage[UTF8]{ctex}"
      \inst{1,2,3}, Ying D. Liu\inst{1,2,3}, L. P. Chitta\inst{4}, Huidong Hu\inst{1,2}, Xiaowei Zhao\inst{5,6}
   }

   \institute{State Key Laboratory of Space Weather, National Space Science Center, Chinese Academy of Sciences, Beijing 100190, People's Republic of China; {\it rwang@swl.ac.cn}\\
        \and
        {Key Laboratory of Solar Activity and Space Weather, National Space Science Center, Chinese Academy of Sciences, Beijing 100190, People's Republic of China}\\
        \and
        {University of Chinese Academy of Sciences, Beijing 100049, People's Republic of China}\\
        \and
        {Max-Planck-Institut f\"ur Sonnensystemforschung, Justus-von-Liebig-Weg 3, 37077 G\"ottingen, Germany}\\
        \and 
        {Key Laboratory of Space Weather, National Satellite Meteorological Center (National Center for Space Weather), China Meteorological Administration, Beijing
        100081, People's Republic of China}\\
        \and
        {School of Earth and Space Sciences, Peking University, Beijing 100871, People's Republic of China}\\
\vs\no
   {\small Received 20xx month day; accepted 20xx month day}
}

\abstract{An extreme ultraviolet (EUV) close-up view of the Sun offers unprecedented detail of heating events in the solar corona. Enhanced temporal and spatial images obtained by the Solar Orbiter during its first science perihelion enabled us to identify clustered EUV bright tadpoles (CEBTs) occurring near the footpoints of coronal loops. Combining SDO/AIA observations, we determine the altitudes of six distinct CEBTs by stereoscopy, ranging from $\sim$1300 to 3300 km. We then notice a substantial presence of dark, cooler filamentary structures seemingly beneath the CEBTs, displaying periodic up-and-down motions lasting 3 to 5 minutes. This periodic behavior suggests an association of the majority of CEBTs with Type I spicules. Out of the ten selected CEBTs with fast downward velocity, six exhibit corrected velocities close to or exceeding 50 km $s^{-1}$. These velocities notably surpass the typical speeds of Type I spicules. We explore the generation of such velocities. It indicates that due to the previous limited observations of spicules in the EUV wavelengths, they may reveal novel observational features beyond our current understanding. Gaining insights into these features contributes to a better comprehension of small-scale coronal heating dynamics.
\keywords{Sun: activity --- Sun: atmosphere --- Sun: corona --- Sun: transition region
}
}

   \authorrunning{Rui Wang et al. }            %author_head in even pages
   \titlerunning{Clustered Dynamic Extreme-Ultraviolet Bright Tadpoles}  % title_head in odd pages

   \maketitle
%% The author head (on even pages) and the title head (on odd pages) will be
%% automatically extracted from \author{} and \title{}. Whenever the title is too long,
%% you will be asked to supply a shorter one by inserting either \authorrunning{} or
%% \titlerunning{} before \maketitle. Anyway, you can specify your own heads.
%%
%%
%% Note: In the following text body of your manuscript, please note several differences from
%%       other major journals:
%% (1) \subsection{Please Capitalize the First Letter of Each Notional Word in Subsection Title}
%% (2) Please Capitalize the First Letter of Each Notional Word in all tables' captions

%
%________________________________________________ sections below
%
\section{Introduction}           %% first-level sections will be auto-capitalized
\label{sect:intro}

With the continuous advancement of observational capabilities, there is a growing interest in studying small-scale phenomena occurring on the Sun, including small-scale coronal jets, mini flares and coronal mass ejections (CMEs), spicules, and the recently widely-discussed campfire events, as well as other small-scale eruptive phenomena and coronal features at a smaller scale \citep{1992Shibata,2007DePontieu,2012Zhangyz,2015Sterling,2021Berghmans,2021Panesar,2023Berghmans}. The increasing attention on these small-scale, and even micro-scale, solar activities is a direct result of the enhanced observational capabilities provided by missions like Solar Orbiter (SolO; \citealt{2020Muller}), enabling for closer observations at distances as close as one-third of the Sun-Earth distance, thereby enhancing our understanding of these phenomena. Furthermore, such small solar activities are closely linked to the major unresolved problem in solar physics of coronal heating, namely how the solar corona reaches temperatures of several million degrees while the photosphere, situated just over a small temperature transition region, remains at approximately 6000 K. This highly temperature-variable region exhibits a significant temperature gradient and typically encompasses the chromosphere, transition region (TR), and lower corona. It is believed that within this narrow region, a significant release of heat and energy occurs, with magnetic field reconnection and various dissipation processes playing crucial roles. Due to observational limitations, these energy release processes often appear as bright spikes or dots in the extreme ultraviolet (EUV) wavelength range. These brignternings exhibit different characteristics depending on their formation mechanisms, which can be inferred from observational features. On the other hand, not all bright structures are necessarily associated with coronal heating. It is important to discern these characteristics based on observations.

Coronal jets, first discovered in coronal X-ray images from Yohkoh \citep{1992Shibata}, are initially thought to be produced by closed-field emergences, following the widely accepted emerging-flux model (e.g., \citealt{1995Yokoyama,2013Archontis,2014Fangf}). However, recent studies suggest a probable association between coronal jets and small-scale filament eruptions \citep{2015Sterling,2023Rui}. Coronal jets occur throughout the Sun in coronal holes, quiet Sun regions, and plages of active region (AR). Solar campfires resembling small-scale jets, manifest as short-lived, EUV brightennings and exhibit diverse complex structures \citep{2021Houzy,2021Panesar,2023Panesar}. They are primarily rooted at chromospheric network boundaries, positioned at heights between 1000 and 5000 km above the photosphere \citep{2021Berghmans}, and may be caused by component magnetic reconnection \citep{2021Chenyj,2024Yurchyshyn}.

An increasing number of observations reveal that solar spicules also exhibit coronal brightening phenomena in the EUV wavelength range (e.g., \citealt{2011DePontieu,2014Pereira,2019Samanta}). \cite{2007DePontieu} identified two types of spicules, i.e., Type I and Type II, in the Ca II H spectral line. They noted that the two types of spicules primarily differ in terms of the speed at which they transport material through the chromosphere and their lifetimes. Type I spicules have speeds of 10 - 40 km $s^{-1}$ and exhibit periodic up-and-down motion with a duration of 3 - 7 minute, while Type II spicules have faster speeds of 50 - 150 km $s^{-1}$ and shorter lifetimes ranging from 10 to 150 s. In the Ca II H filtergrams, Type II spicules fade away, while Type I spicules commonly manifest as parabolic space-time diagrams. The formation mechanism of Type I spicules is believed to be associated with the leakage of photospheric oscillations and convective motions into the chromosphere along magnetic flux concentrations, where shock waves form and propel plasma jets upwards. On the other hand, Type II spicules are thought to be related to magnetic reconnection and are associated with coronal heating. The leading edge of Type II spicules has been observed to brighten in the EUV 171 \AA~and can be tracked to higher altitudes compared to the Ca II H line, indicating plasma heating to coronal temperatures of at least 1 to 2 MK \citep{2011DePontieu}. \cite{2014Pereira} observed similar parabolic paths in Type II spicules using the EUV 304 \AA~wavelength and the Mg II and Si IV spectral lines of the IRIS instrument while fading in the Ca II H line, suggesting associated plasma heating. \cite{2019Samanta} found that spicules are related to opposite-polarity magnetic flux and observed EUV brightenings at the top of the spicules, supporting evidence of plasma heating to coronal temperatures. They observed heated material with enhanced brightening at the top of spicules, and sometimes, falling back from the corona. However, their work did not explicitly differentiate between Type I and Type II spicules.

Another interesting phenomenon that can cause EUV brightenings, but is often overlooked, is the phenomenon associated with coronal rain. In a recent study, \cite{2023Antolin} discovered a phenomenon similar to the fireball phenomenon on Earth, which is linked to meteoric ablation. During the rapid descent, the coronal rain clumps experience compression and friction with the surrounding solar atmosphere, leading to their heating. Moreover, the rebound of the rain clumps after the fall also contributes to brightenings observed in the EUV wavelength range. \cite{2022Antolin} and \cite{2022Lixh} both analyzed the pressure and temperature surrounding the rain clumps when they reach chromospheric heights, and they found the formation of high-temperature and high-pressure regions around the clumps. Observations indicate that the descent velocity of this coronal rain clumps can reach speeds of 70 - 80 km $s^{-1}$ \citep{2023Antolin}. \cite{2022Lixh} also presented synthesized EUV 171 \AA~images, unveiling brightenings observed around the coronal rain clumps.

There are other phenomena, such as jetlets observed at the base of solar coronal plumes \citep{2022Kumar} and sparkling EUV bright dots observed with the Hi-C instrument \citep{2014Regnier}, dynamic fibrils at the coronal base \citep{2023Mandala,2023mandalb}. These brightenings within the EUV wavelength range appear to indicate higher coronal temperatures, with the majority of them potentially playing a significant role in coronal heating. In our study, utilizing high-resolution EUV observations from the SolO mission, we have also uncovered some interesting small-scale phenomena that have rarely been reported in previous studies. In the next section, we will present our observational findings. Finally, in the last section, we will compare our discoveries with previous observations and explore possible mechanisms underlying these phenomena.

\section{Observations and Anaylsis}
\label{sect:Obs}

SolO's perihelion, its closest approach to the Sun, took place on 2022 March 26. The spacecraft was inside Mercury's orbit, approximately one-third of the distance between the Sun and Earth. From this vantage point, the EUV High Resolution Imager (HRI$_{EUV}$) of the Extreme Ultraviolet Imager (EUI; \citealt{2020Rochus}) provides continuous high resolution observations with a 3-second cadence from 00:03:00 UT to 00:47:57 UT on March 30. The angular pixel size is 0$^{\prime\prime}$.492, corresponding to 118 km on the solar surface. HRI$_{EUV}$ observed a cluster of dynamic EUV bright tadpole-like structures near the footpoints of corona Loops. Figure 1a illustrates SolO's near-quadrature position with Earth on 2022 March 30, at an approximate separation angle of $\sim$60$^{\circ}$ from the AR. The animation of the zoomed insert of Figure 1b reveals that these clustered EUV bright tadpoles (CEBTs, hereafter) are very dynamic and exhibt two-direction movements, i.e., upward and downward. The Atmospheric Imaging Assembly (AIA; \citealt{2012Lemen}) on board the Solar Dynamics Observatory (SDO; \citealt{2012Pesnell}) also captured these bright features. However, the AIA images appear mosaic-like, displaying sporadic discernible brightening structures. Nonetheless, by using nearly simultaneous observations from two different angles, we can employ images from AIA and HRI$_{EUV}$ for stereoscopic measurements, enabling us to estimate the heights of the CEBTs. 

%%%%%%%%%%%%%%%%%%%%%%%%%%%%%%%%%%%%%%%%%%%%%%%%%%%%%%%%%%%%%%
%%     Examples for figures using graphicx for LaTeX 2e
%%               -- our recommended way for embodying graphics
%%%%%%%%%%%%%%%%%%%%%%%%%%%%%%%%%%%%%%%%%%%%%%%%%%%%%%%%%%%%%%
%
%      A figure as large as the width of the column
%-------------------------------------------------------------
\begin{figure}
   \centering
   \includegraphics[width=1.0\textwidth, angle=0]{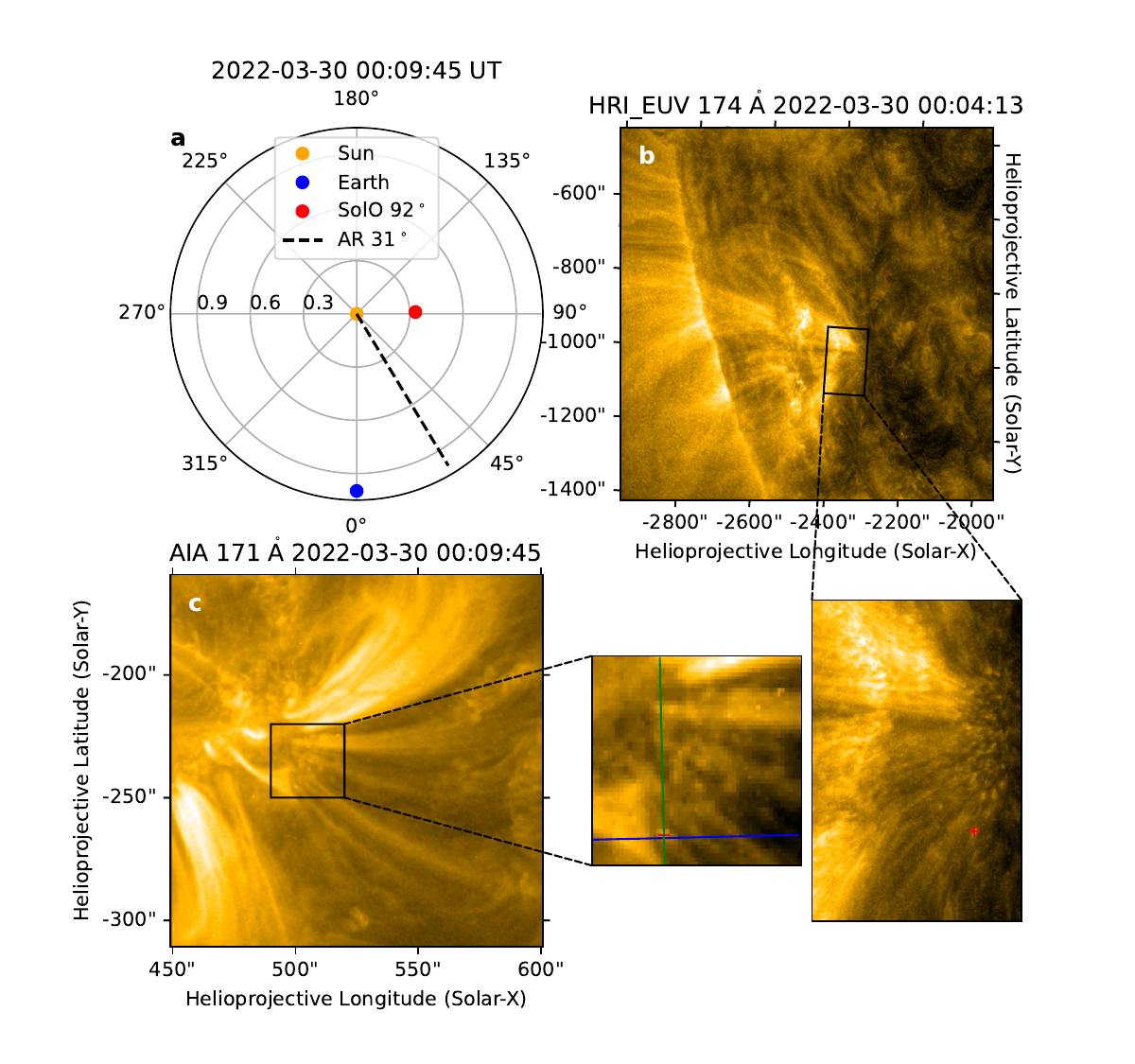}
   \caption{Stereoscopic measurements using SDO and SolO. (a) Heliospheric positions of planets and spacecraft. (b) and (c) CEBTs observed from EUI/HRI$_{EUV}$ and AIA views with localized zoomed inserts, respectively. The red crosses and the intersecting green and blue lines in the inserts represent stereoscopic measurements using the two instruments. Animation of the inserts is available.}
   \label{Fig1}
   \end{figure}
%
%      One column rotated figure
%-------------------------------------------------------------
We employ stereoscopic observation on CEBTs identifiable in both the HRI$_{EUV}$ 17.4 nm and AIA 17.1 nm images. This method involves identifying the same feature (pixel) in both images and calculating the lines of sight (LOS) passing through the corresponding pixels. This enables us to determine the position of the intersection point in 3D space (more details on methodology please refer to Appendix A). To account for the time difference between the two images, we consider the light travel time difference (about 330 s) between the Sun to each instrument. After correcting for light travel time, we use HRI$_{EUV}$'s higher temporal resolution images (3 s) to match the lower temporal resolution images of AIA (12 s), reducing the time difference between the two images to within 2 s.
%----------------------------------------------------------------------
\begin{figure}
   \centering
   \includegraphics[width=1.0\textwidth, angle=0]{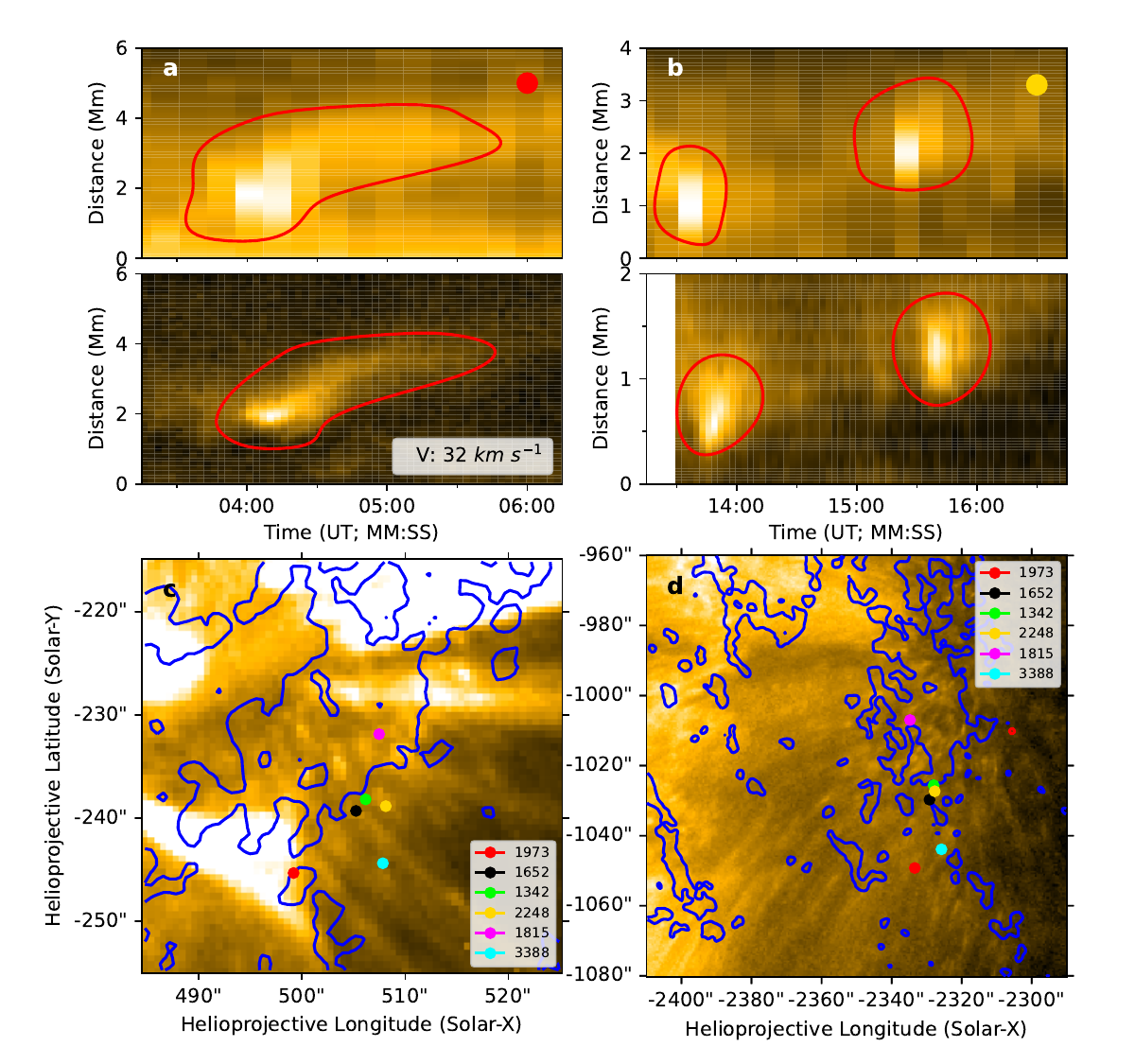}
   \caption{Two typical examples of upward-moving CEBTs (U-CEBTs) observed from two perspectives (the other four in the Figure A.1). Panels (a) and (b) show time-distance plots of two typical U-CEBT events from AIA (top) and EUI/HRI$_{EUV}$ (bottom) using radially outward slits along the loop structures. The times are the SolO time. (c) and (d) Six U-CEBTs, represented by colored dots, are visually consolidated into a single image obtained from SDO and SolO, respectively. Their altitudes, determined through stereoscopy, are annotated. The corresponding LOS magnetograms from HMI and PHI/HRT are overlaid on the EUV images.}
   \label{Fig2}
   \end{figure}

%%%%%%%%%%%%%%%%%%%%%%%%%%%%%%%%%%%%%%%%%%%%
We identify six CEBTs by comparing the images from the AIA and HRI$_{EUV}$. The morphological evolution of these six points in the two images from different instruments, as depicted in the upper and lower panels of Figure 2a and 2b, exhibits overall consistency. Furthermore, these bright points need to exhibit relatively high brightness, enabling their identification in both images. Moreover, these points should appear relatively isolated in the HRI$_{EUV}$ image. This is because, compared to the AIA image, the HRI$_{EUV}$ image contains a larger number of bright points, which can introduce interference in our stereoscopic measurements. Therefore, the selected CEBTs are those that exhibit a clear correspondence between the AIA and HRI$_{EUV}$ images. Our results inidcate that the height distribution of these six CEBTs ranges from $\sim$1300 to 3300 km (see Figure 2 and Figure A.1), encompassing the chromosphere, TR, and lower corona. 

We consolidate these six CEBTs from different time frames into a single image after compensating for the effects of differential rotation. This allows us to a direct comparison of their approximate altitudes and relative distribution of these points. We also overlay LOS magnetograms from the Helioseismic and Magnetic Imager (HMI; \citealt{2012Scherrer,2012Schou}) on board SDO and the High Resolution Telescope (HRT) of the Polarimetric and Helioseismic Imager (PHI; \citealt{2020Solanki}) on board SolO onto the AIA and HRI$_{EUV}$ images, respectvely. However, since the L2 data\footnote{We used level 2 (L2) EUI data, which can be accessed via \url{https://www.sidc.be/EUI/data/releases/202301\_release\_6.0}. Information about the data processing can be found in the release notes at DOI: \url{https://doi.org/10.24414/z818-4163}.} of the HRT LOS magnetograms and the HRI$_{EUV}$ images lack limb fitting correction, resulting in telescope pointing deviations caused by thermal elastic deformation \citep{2020Auchere}. As a result, direct image alignment is not possible. To address this issue, we employ a method where we use the HMI magnetogram, corrected for projection effects, to rectify the deviation in the HRT images; and use nearby EUI/Full Sun Imager (FSI) 174 \AA~images to correct the deviation in the HRI$_{EUV}$ images. This method uses a specific feature visible in both FSI (HMI) and HRI$_{EUV}$ (HRT) images to align them by iteratively adjusting the CRVAL values of the HRI$_{EUV}$ (HRT) image\footnote{\url{https://github.com/SolarOrbiterWorkshop/solo8_tutorials/blob/main/EUI_tutorial/3_alignment_b.ipynb}}. Incidentally, we removed the spacecraft jitter in the L2 HRI$_{EUV}$ data using a cross-correlation technique \citep{2022Chitta}.

\begin{figure}
   \centering
   \includegraphics[width=1.0\textwidth, angle=0]{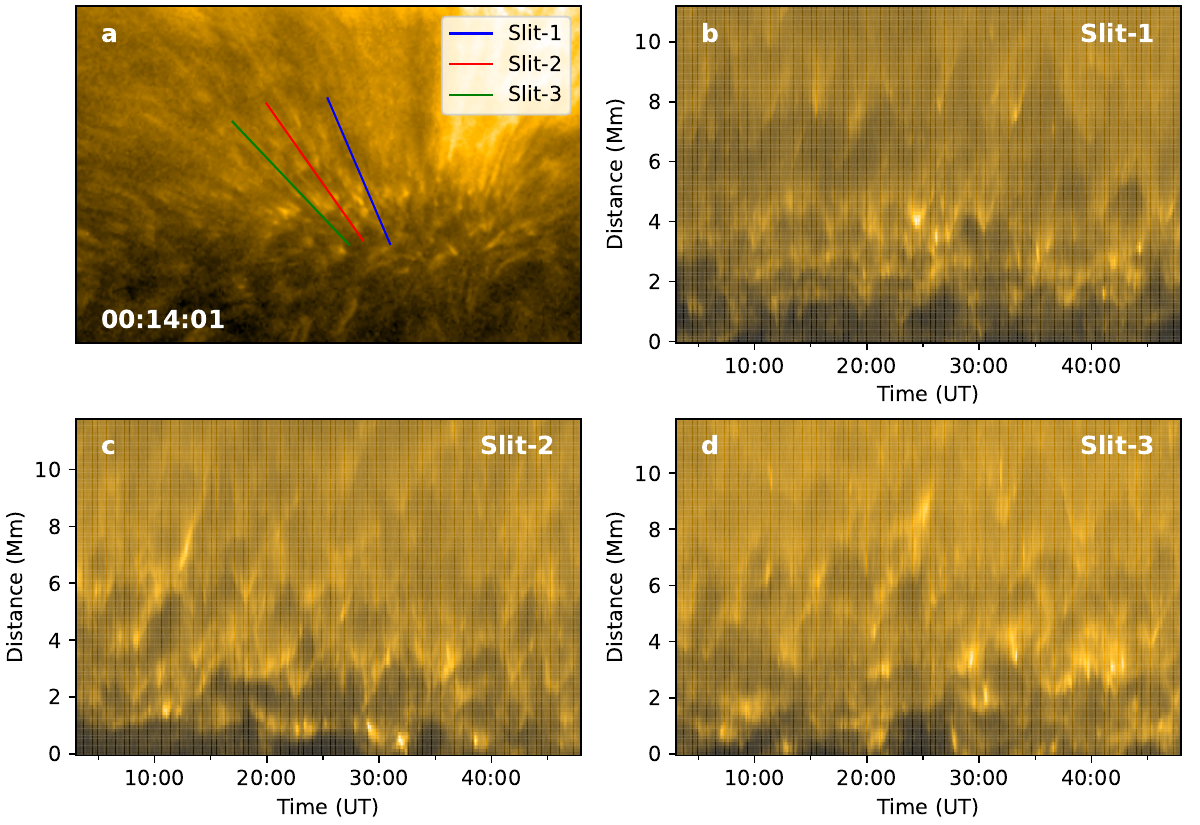}
   \caption{The periodic measurements of dark and cooler filamentary structures associated with CEBTs. (a) Lines in different colors represent the artificial slits used for generating time-distance plots, depicted in panels (b)-(d). The directions of the slits are from bottom right to top left. Animation of (a) is available.}
   \label{Fig3}
\end{figure}

%%%%%%%%%%%%%%%%%%%%%%%%%%%%%%%%%%%%%%%%%

From the perspective of SDO, it is evident that these CEBTs are predominantly located at the plage of the AR. Similarly, when observed from the SolO perspective, the majority of CEBTs are concentrated above and around the relatively strong negative polarity magnetic field. During the observed period, the AR was in a decaying phase without any significant sunspots present. The magnetic field configuration may play a crucial role in the occurrence of CEBTs, which brings to mind the possibility of its association with spicules (or dynamic fibrils; \citealt{2023Mandala,2023mandalb}). Type I spicules are more likely to manifest around ARs. \cite{2007DePontieu} noted that type I spicules exhibit a motion along a parabolic path, i.e., ejected then fall, while type II spicules tend to fade away after rapid ejection. However, recent studies have revealed a more complex situation. \cite{2014Pereira} observed that both types of spicules can show parabolic trajectories in multiple spectral lines of the IRIS instrument. \cite{2015Skogsrud} further discovered that Ca II spicules significantly weaken, with their descending phase primarily visible in IRIS and AIA data, while in Ca II data it can only be revealed through aggressive filtering. These findings indicate that spicule dynamics exhibit complex characteristics across different spectral lines. Our observations are based on the 17.4 nm EUV images, which can capture higher temperature phenomena than the observations in the calcium line. By careful examiation, we notice (see animation of Figure 3a) that compared to the slower upward motion, the downward motion of CEBTs appears to dominate, and the motion speed seems faster. Hereafter, the abbreviations U-CEBTs and D-CEBTs will be employed to represent upward-moving and downward-moving CEBTs, respectively.

\cite{2011DePontieu} and \cite{2019Samanta} reported the presence of EUV brightenings at the tops of Type II spicules observed at 171 \AA. These brightenings at the tops of Type II spicules bear some resemblance to the CEBTs we have observed. However, when considering the U-CEBTs (see Figure 2 and Figure A.1), their velocities (corrected for projection effects; for details, refer to the velocity correction of D-CEBTs below) are generally much lower than the typical speeds of 50 - 100 km$s^{-1}$ observed for Type II spicules. In terms of upward motion velocity, this type of motion is more similar to dynamic fibrils \citep{2023Mandala}. Figure 3a reveals the presence of dark, cooler filamentary structures (FS) in addition to the bright CEBTs. Due to the projection effect and the dimness of the background, it is challenging to determine the spatial relationship between these features and the CEBTs in most cases. We place the slits in the areas where CEBTs are concentrated to measure the movement characteristics of these FS. The time-distance evolution plots in Figures 3b-d indicate that the majority of FS are undergoing periodic oscillations. However, FS in front or behind can impact periodicity measurements due to on-disk observations. Therefore, we select three slits to characterize the existence of this periodicity through measurements at different locations. A rough estimate suggests an oscillation period ranging from 3 to 5 minutes. The CEBTs correspond to bright structures at the tips of the FS. Based on current results, most CEBTs are related to FS with clear periodicity, while others are associated with FS lacking obvious periodic behavior. Overall, periodic FS predominate in this area.

\begin{figure}
   \centering
   \includegraphics[width=1.0\textwidth, angle=0]{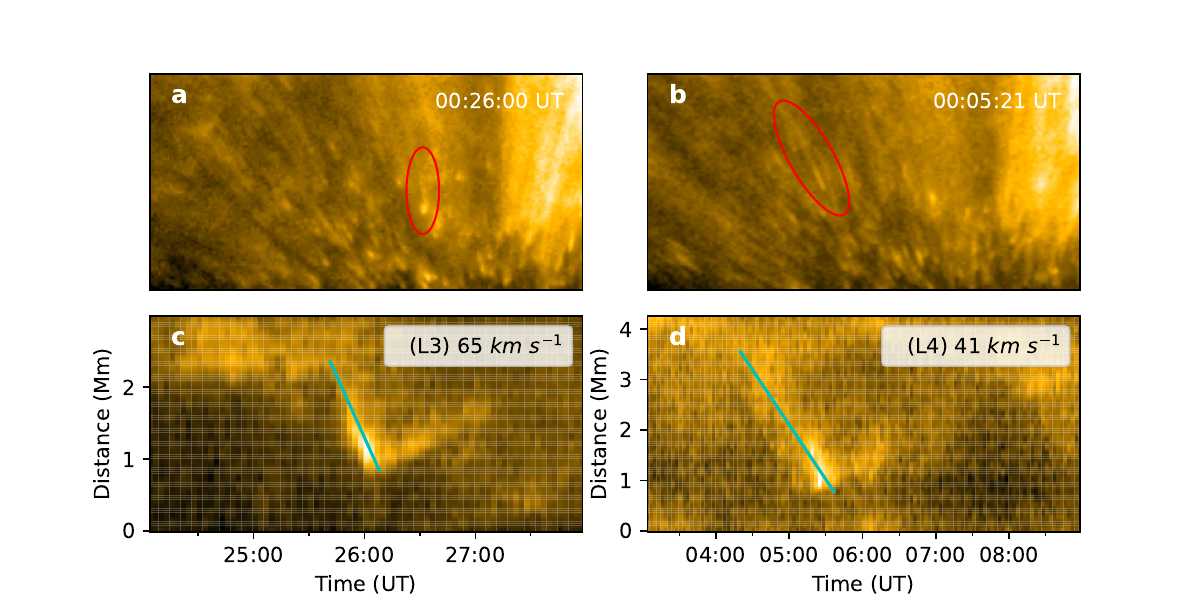}
   \caption{(a) and (b) Typical examples of downward-moving CEBTs (D-CEBTs) circled by red eclipses (numbered as 3 and 4 in Figure 5). The times are the SolO time. (c) and (d) Corresponding time-distance plots of D-CEBTs, with downward-moving velocities annotated by cyan lines.}
   \label{Fig4}
\end{figure}
%%%%%%%%%%%%%%%%%%%%%%%%%%%%%%%%%%%%%5
Figure 4 presents the corrected radial velocities of two typical D-CEBTs. We have accounted for the projection effect, facilitating better velocity comparison with spicules previously observed at the limb. We assume that the velocity is along the radial direction. In the heliocentric-radial coordinate system, V$_{rad}$= V$_{obs}$ / ($\rho$/R$_\odot$), where $\rho$ is the impact parameter, and R$_\odot$ is the solar radius. The D-CEBTs in Figure 4a reaches a velocity of 65 km $s^{-1}$, surpassing the typical upward velocity of Type I spicules. Additionally, both D-CEBTs in Figure 4 exhibit a distinct rebound motion (as shown in Figure 4c and 4d). Figure A.2 and A.3 provide the velocity measurements for the remaining eight D-CEBTs we selected. More than half of these CEBTs have velocities of around 50 km $s^{-1}$ or higher. Figure 5 shows the consolidation of these ten D-CEBTs into a single image, showing the relative distribution of these D-CEBTs. We are unable to estimate the heights of these CEBTs since we do not observe distinct D-CEBTs in the AIA. However, from Figure 5, we can make a rough judgment, for example, that D-CEBT 2 and 4 do not occur at higher altitudes but rather appear as a visual effect due to projection. In reality, they likely occur in the lower solar atmosphere at a relatively distant location.
%-----------------------------------
\begin{figure}
   \centering
   \includegraphics[width=1.0\textwidth, angle=0]{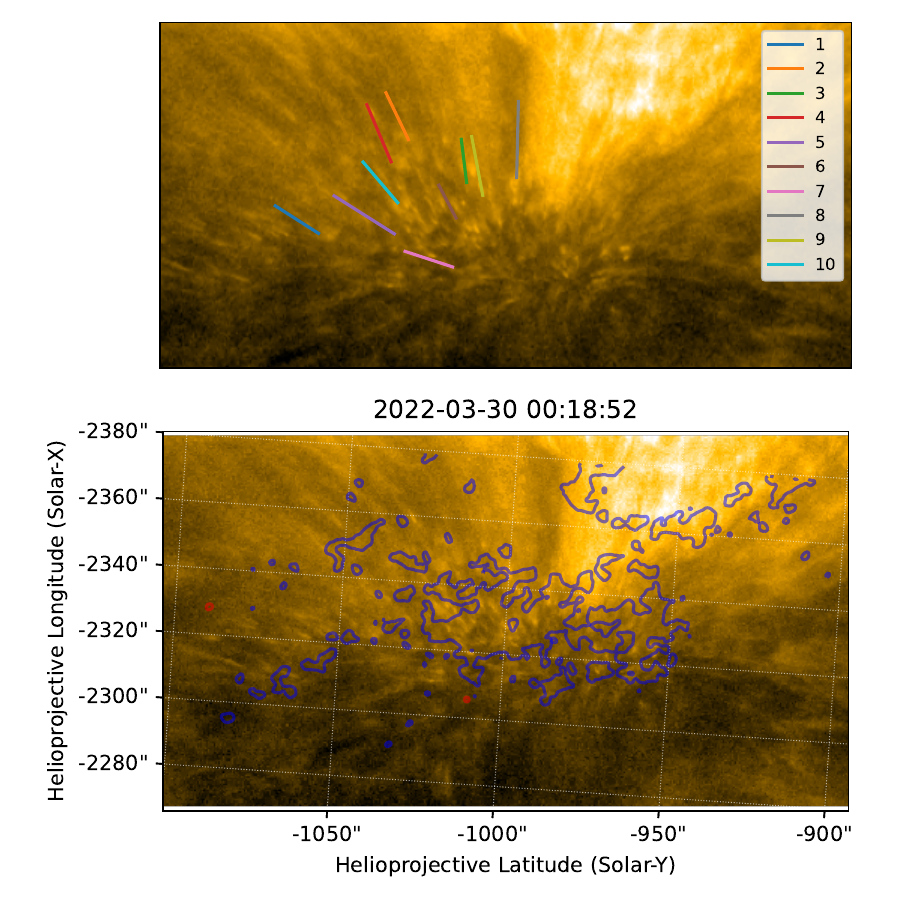}
   \caption{Top: ten examples of D-CEBTs numbered with different colored lines consolidated into a single HRI$_{EUV}$ image. Bottom: corresponding coordinated image overlaid with PHI/HRT magnetogram.}
   \label{Fig5}
\end{figure}
%%%%%%%%%%%%%%%%%%%%%%%%%%%%%%%%

Figure 6a displays the distribution of the PHI/HRT LOS magnetogram with a threshold of $\pm$100 G. In contrast to the LOS magnetogram obtained from HMI, we observe a higher occurrence of positive polarity magnetic fields (indicated by red color) in the weak-field region. These positive polarity fields emerge around the prevailing negative polarity fields and undergo a continuous emergence-cancellation process \citep{2022Rui}, which may result in magnetic transients transferring the magnetic energy to power the higer corona \citep{2019Chitta}. The HRI$_{EUV}$ (Figure 6b) and the Spectral Imaging of the Coronal Environment (SPICE; \citealt{2020Anderson}) images (Figure 6c-6h) on board the SolO are employed to aid in the analysis of the temperature distribution of the background plasma in the source region. The area above and surrounding the negative polarity magnetic fields exhibits a broader range of plasma temperatures (10$^4$ - 10$^6$ K). However, in the region slightly further away from the weak magnetic field, the plasma density and temperature are relatively lower. This is probably the reason why we can observe downward motions similar to D-CEBT 2 and 4 occurring in those areas.
\begin{figure}
   \centering
   \includegraphics[width=1.0\textwidth, angle=0]{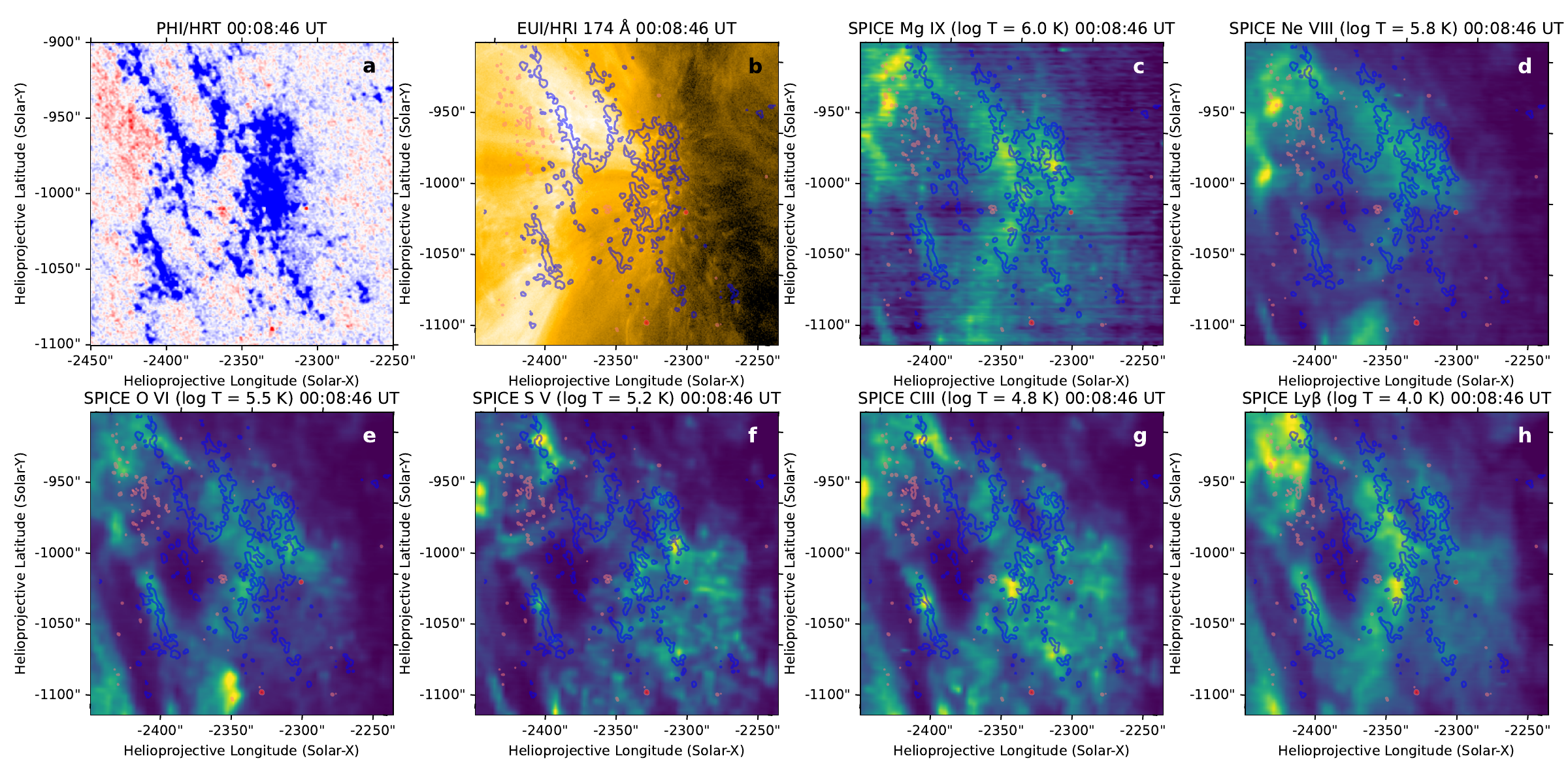}
   \caption{Joint observations of Multi-instruments from SolO. (a) PHI/HRT LOS magnetogram. Positive (red) and negative (blue) magnetic fields scaled to $\pm$100 G. (b) and (c)-(h) Co-aligned EUI/HRI$_{EUV}$ image and SPICE images in different spectral lines, both overlaid with the reprojected HRT magnetograms.}
   \label{Fig6}
\end{figure}

\section{Summary and Discussion}
\label{sect:discussion}

Due to its unprecedented proximity to the Sun, the EUI instrument is capable to capture highly detailed images of the solar atmosphere. These images have revealed fascinating small-scale features known as CEBTs, referring to a cluster of dynamic EUV bright tadpole-like structures located near the footpoints of coronal loops. Through stereoscopic observations, it has been observed that some of CEBTs are distributed at heights ranging from $\sim$1300 to 3300 km, typically belonging to altitudes of the chomosphere to the lower corona, with upward velocities of only a few tens of kilometers per second. In constrast, certain downward CEBTs have been detected with velocities reaching or even exceeding 50 km $s^{-1}$. Although classified as small-scale features, CEBTs span a remarkable distance of approximately 25000 km across the Sun, roughly twice the diameter of the Earth. In this region, there are a multitude of hot CEBTs and colder gas extending in various directions. It is most likely that the CEBTs are located at the top of this colder gas. Analysis of the typical features has revealed that the colder gas (or filamentary structures) undergoes oscillations with periods lasting from 3 to 5 minutes. Based on these observations, it is believed that the majority of CEBTs probably represent bright structures situated at the uppermost part of Type I spicules.

\cite{2007DePontieu} mentioned the relationship between the formation of Type I spicules and the inclination angle of magnetic fields. They indicated that Type I spicules are generated when the energy released from oscillations and convective motions beneath the solar photosphere creates shock waves through magnetic flux tubes, propelling plasma upwards along these tubes. When the magnetic inclination angle is appropriate, such as in the vicinity of ARs, there is an increased occurrence of leakage of oscillations within the 3 - 7 minute range. This leads to the formation of more energetic Type I spicules compared to those found in coronal holes. Consequently, this may explain the detection of a higher number of CEBTs around the footpoints of coronal floops in dispersed active regions

Interestingly, a prominent characteristic of these CEBTs is their predominant downward motion. Out of the ten selected CEBTs with fast downward velocities, six demonstrate corrected speeds around or exceeding 50 km $s^{-1}$, typically associated with the ejection velocity of Type II spicules. However, previous studies demonstrated that Type II spicules fade in the Ca II H spectral line. Although similar parabolic trajectories are observed in Type II spicules in the EUV 304 \AA~wavelength and the spectral lines of Mg II and Si IV, limited evidence exists regarding the rapid downward movements of Type II spicules, especially in the EUV 17.1 nm wavelength. Several possible explanations are considered for the generation of these fast downward CEBTs. One possibility is that they originate from magnetic reconnection by fields braiding from higher altitudes, giving the appearance of downward motion because of projection effects, or from the ejections of micro-scale jets resembling small ``campfires'' which appear to have downward motion due to projection effects or descend along a small loop's descending half-loop. Regarding the former, from the morphology of the CEBTs in the animation of Figure 3a or Figure 4, it is observed that the tapered tails of the CEBTs are located above. If they were outflows generated by magnetic reconnection from higher locations, they would have the tapered tails pointing downward. As for the latter, no low-lying coronal loop structures have been observed in that region, and obtaining loop structures through extrapolation of the magnetic field has failed. Thus, they should not be coronal jets.

Another potential explanation is that they are attributed to micro-scale coronal rains, potentially the smallest structures of this kind observed so far. In the same AR, Antolin et al. (2023) observed a fireball-like phenomenon resembling meteoric ablation on Earth. However, distinct observational features of coronal rain, namely, the presence of cool and dense plasma clumps, have not been observed in our case. Instead, only detected bright spikes without any indication of dark clumps have been detected.

%%%%%%%%%%%%%%%%%%%%%%%%%%%%%%%%
\begin{figure}
   \centering
   \includegraphics[width=1.0\textwidth, angle=0]{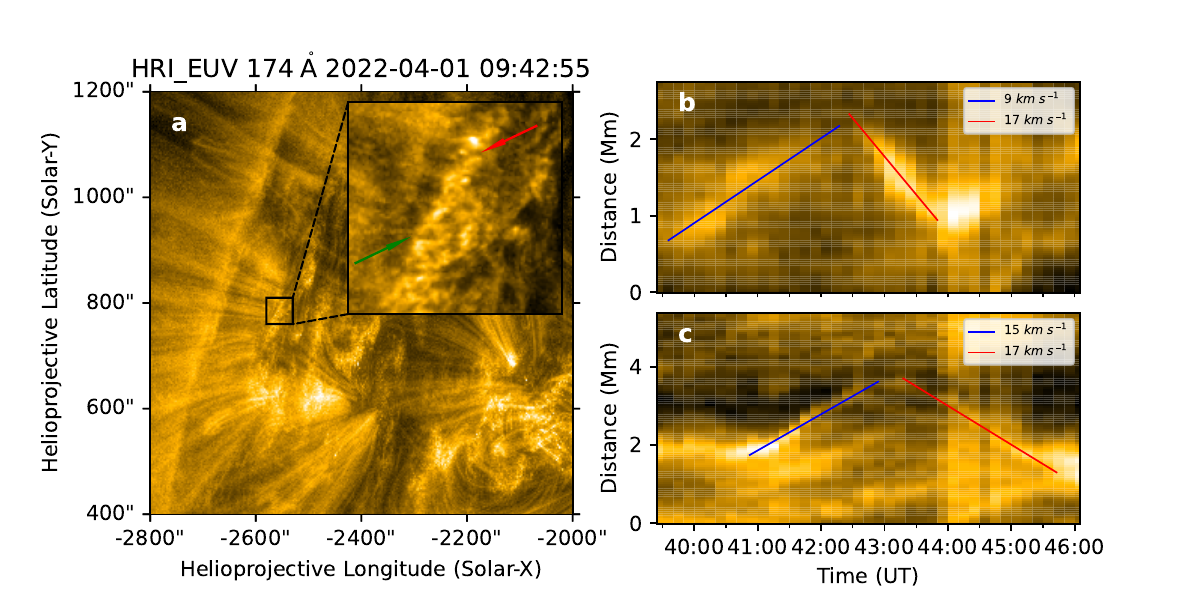}
   \caption{(a) CEBTs occured on 2022 April 1. The localized zoomed inserts show two obvious Type I spicules near the footpoints of coronal loops. Animation is available. (b) and (c) Time-distance plots corresponding to the spicules indicated by red and green arrows, respectively.}
   \label{Fig7}
\end{figure}
%%%%%%%%%%%%%%%%%%%%%%%%%%%%%%%%%%%%%%%%%%

If we assume that these D-CEBTs belong to the descending phase of Type I spicules, it raises the question as to why the corresponding ascending phase has not been observed. In fact, for the selected 10 D-CEBTs, we cannot definitively identify the corresponding U-CEBTs. One reason could be the large number of overlapping CEBTs and dark filamentary structures, making it difficult to distinguish individual events. Another possibility is that their ascending phases do not exhibit distinct brightening features. As shown in Figure 3, brightening is observed intermittently and is not consistently present. Now, we raise the question as to why some D-CEBTs exhibit such high velocities, which leads us to wonder if their ascending speeds are similarly fast. Analyzing the event that took place on 2022 April 1, Figure 7a demonstrates the presence of similar CEBT structures around the negative polarity of the AR in the decaying phase. In the zoomed-in image of the footpoints in the coronal loop, we identified two prominent spicule structures undergoing up-and-down movements (indicated by the red and green arrows). Surprisingly, Figure 7b shows that the brightening structure at the top of the spicule (indicated by the red arrow) exhibits a downward velocity almost twice that of the upward velocity. Furthermore, the brightening structure exhibits a higher brightness in the downflow compared to the upflow. However, this is not a universal trend, as Figure 7c shows comparable upward and downward velocities for the spicule. Although the observed D-CEBTs in this particular event did not display such high velocities, it is possible that if a spicule has a low upward velocity below 50 km s$^{-1}$, it could have a downward velocity exceeding 50 km s$^{-1}$. This, to some extent, can explain the high velocities observed in our D-CEBTs. Similar phenomena known as dynamic fibrils have been reported by \cite{2023Mandala,2023mandalb}. These studies suggest that these dynamic fibrils are shock-driven chromospheric phenomena with EUV brightenings as the tips of the dynamic fibrils. They provided evidence connecting the EUV dynamic fibrils to their chromospheric counterparts, highlighting that the dynamic fibrils reflect coronal temperatures and exhibit an intensity evolution similar to type I spicules.

In fact, previous studies did not observe or paid much attention to downflow motions in the EUV 171 or 174 \AA~wavelength range. For instance, \cite{2011DePontieu} demonstrated the upflow of Type II spicules in He II 304 \AA, Fe IX 171 \AA, and Fe XIV 211 \AA, but did not indicate the presence of downflows at these wavelengths. \cite{2014Pereira} observed similar parabolic trajectories in Type II spicules using the EUV 304 \AA~wavelength and the spectral lines of Mg II and Si IV, which correspond to relatively lower temperatures compared to 171 or 174 \AA. \cite{2019Samanta} presented heated material with enhanced brightness at the top of spicules at 171 \AA, occasionally falling back from the corona. However, their study did not explicitly differentiate between Type I and Type II spicules. Indeed, this rapid descending flow may also result from the descent of Type II spicules intermingled within Type I spicules. In Figure 3d, we observe sporadic bright thread-like features that are not related to the filamentary structures with clear periodicity. It is worth noting that research on the descending motion of spicules in this specific wavelength range remains limited. In future studies, we intend to further investigate the underlying mechanisms driving this rapid descending flow.

In conclusion, we have observed some previously unnoticed small-scale coronal features in the EUV 174 \AA~wavelength range. We believe that the majority of CEBTs are brightening structures associated with Type I spicules. Apart from the events we have mentioned, similar CEBT structures can be seen in many high-resolution images from the EUI. While it remains uncertain whether these heating processes associated with the bright structures contribute to coronal heating or if they are merely spurious elements that confound our understanding of coronal heating. Understanding these intricate EUV brightening structures will enhance our knowledge of features in the lower solar atmosphere. With the future release of more high-resolution images from SolO, we will be able to further explore the physical nature of these fine structures.

\normalem
\begin{acknowledgements}
The research was supported by National Key R\&D Program of China No. 2022YFF0503800 and No. 2021YFA0718600, the Strategic Priority Research Program of the Chinese Academy of Sciences (NO. XDB0560000), NSFC under grants 12073032, 42274201, 42150105, and 42204176, and the Specialized Research Fund for State Key Laboratories of China. We acknowledge the use of data from Solar Orbiter and SDO. Solar Orbiter is a space mission of international collaboration between ESA and NASA, operated by ESA. The EUI instrument was built by CSL, IAS, MPS, MSSL/UCL, PMOD/WRC, ROB, LCF/IO with funding from the Belgian Federal Science Policy Office (BELSPO/PRODEX PEA 4000134088); the Centre National d'Etudes Spatiales (CNES); the UK Space Agency (UKSA); the Bundesministerium f\"{u}r Wirtschaft und Energie (BMWi) through the Deutsches Zentrum f\"{u}r Luft- und Raumfahrt (DLR); and the Swiss Space Office (SSO). L.P.C. gratefully acknowledges funding by the European Union (ERC, ORIGIN, 101039844). Views and opinions expressed are however those of the author(s) only and do not necessarily reflect those of the European Union or the European Research Council. Neither the European Union nor the granting authority can be held responsible for them.

\end{acknowledgements}

\bibliography{references}{}
\bibliographystyle{raa}

%%%%%%%%%%%%%%%%%%%%%%%%%%%%%%%%%%%%%
\appendix    
\section{Supplementary Examples of U-CEBTs and D-CEBTs}
A stereoscopy method is used to estimate the heights of CEBTs. We implement a stereoscopy method using a Python program, with the code based on the SolarSoft routine ``scc\_measure.pro'' written in the IDL language. We first locate isolated CEBT in an EUI image. Then, we draw a line (blue line in the zoomed insert of Figure 1c) perpendicular to the image plane through the CEBT. We convert this line's coordinates from Helioprojective Cartesian to Heliocentric Cartesian (HCC), an observer-based coordinate system. Next, we transform these coordinates to Heliographic Earth Equatorial (HEEQ), an Earth-referenced coordinate system. In the AIA images, we draw another line (green line) through the corresponding CEBT in HCC coordinates, perpendicular to the transformed line from EUI. We then find the intersection point of these two lines. This intersection point is converted back to HEEQ coordinates, which provide its latitude, longitude, and radial height. By subtracting the solar radius from this radial height, we obtain the point's altitude. The height of the intersection point represents the altitude of the CEBT feature.

Since we only measure the bright points from two angles, we do not define uncertainties as done in \cite{2006Inhester}. One AIA pixel corresponds to 441 km on the solar surface, and one HRI$_{EUV}$ pixel corresponds to 118 km. The error is then taken as the maximum of 441 km if a CEBT can be localized within one pixel in both the HRI$_{EUV}$ and AIA images. However, as pointed by \cite{2021Berghmans}, the requirement of four HRI$_{EUV}$ pixels and two AIA pixels may be necessary to distinguish a complete structure. Therefore, this error should be accounted for as 882 km. This represents the random error associated with the limited resolution of the telescopes. Additionally, we employ a time-series-based morphological comparison, as shown in Figure 2a, where the selected six CEBTs all exhibit clearly corresponding morphological features in the images from both instruments. We can confidently ascertain that the bright spikes observed in both AIA and HRI$_{EUV}$ images originate from the same CEBTs. Furthermore, we use the method mentioned in Section \ref{sect:Obs} to address the limb fitting issue in the L2 data of HRI$_{EUV}$ and HRT. As a result, our estimation of CEBT heights should be relatively accurate.

\begin{figure}
   \centering
   \includegraphics[width=1.0\textwidth, angle=0]{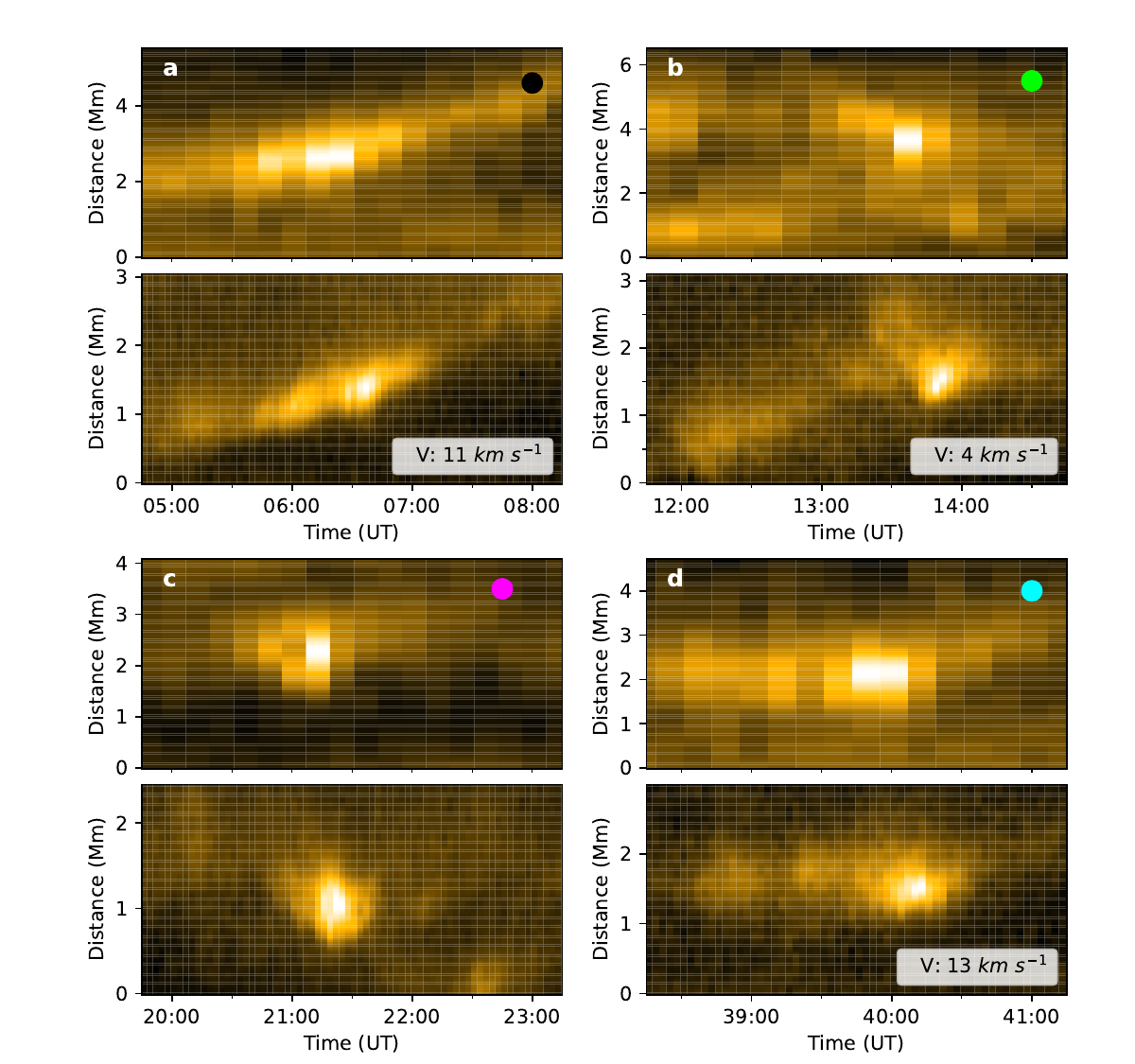}
   \caption{Same as (a) and (b) in Figure 2.}
   \label{Fig.A1}
\end{figure}

%-------------------------------
\begin{figure}
   \centering
   \includegraphics[width=1.0\textwidth, angle=0]{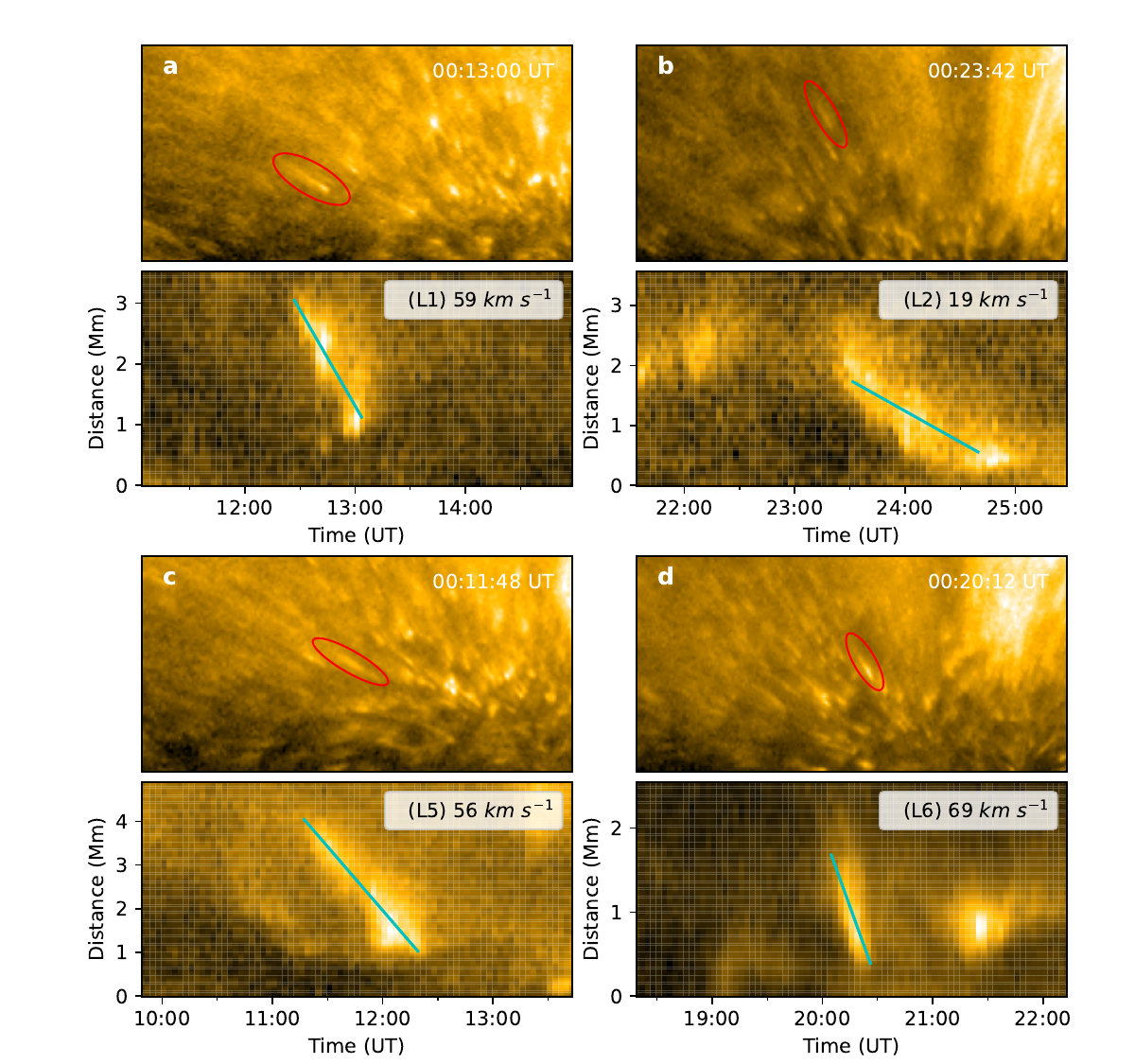}
   \caption{Same as in Figure 4, but for D-CEBTs of 1, 2, 5, 6.}
   \label{Fig.A2}
\end{figure}
%-------------------------------

\begin{figure}
   \centering
   \includegraphics[width=1.0\textwidth, angle=0]{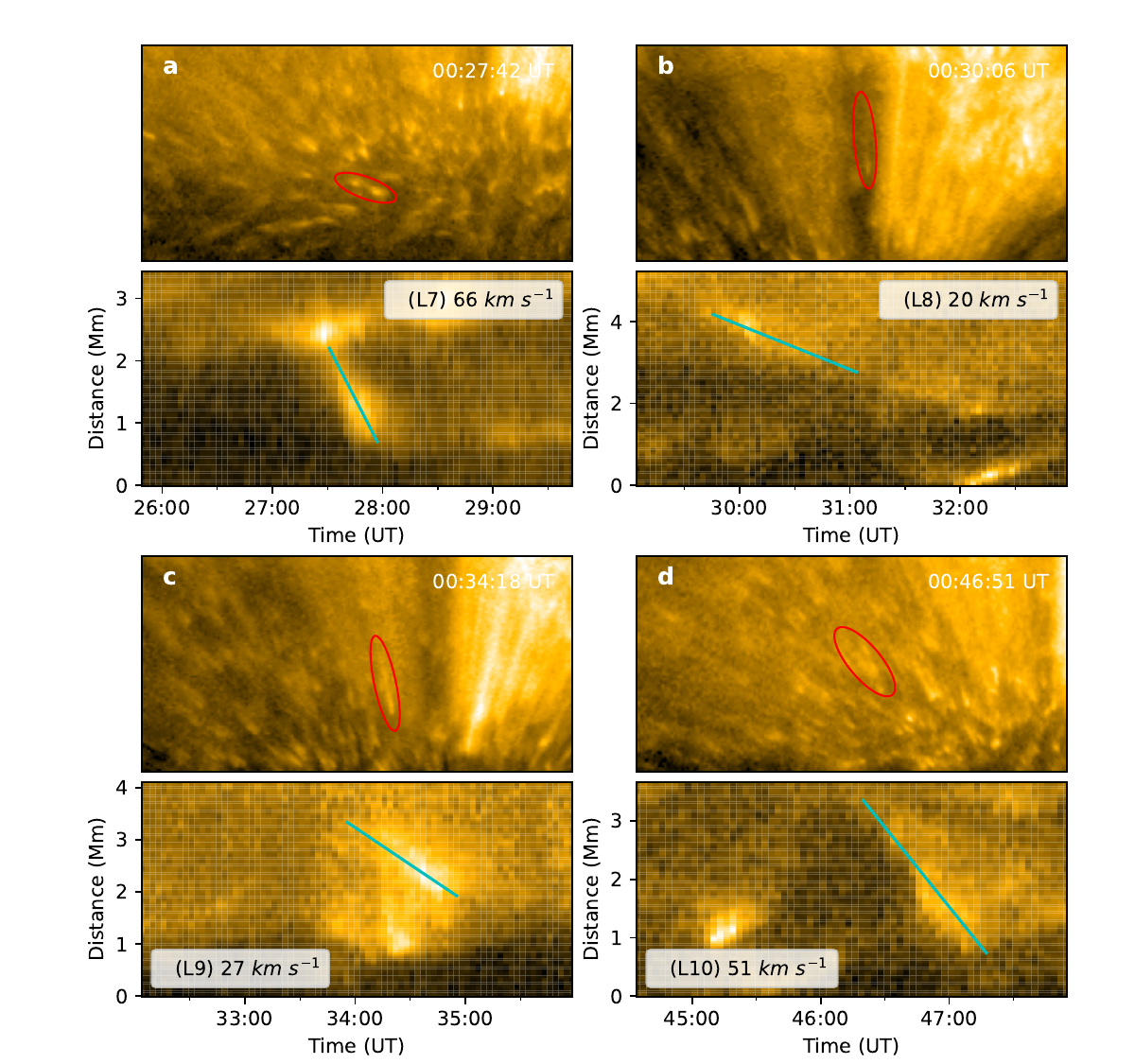}
   \caption{Same as in Figure 4, but for D-CEBTs of 7, 8, 9, 10.}
   \label{Fig.A3}
\end{figure}

%\begin{thebibliography}{99}
%% you can type \apj for ApJ, \aap for A&A, \apss for Ap&SS, etc. Please consult
%% the macro chjaa.cls. You can also find them in aasguide.tex (AASTeX for ApJ, AJ, PASP)
%% Please follow the format of ChJAA's reference list

%\end{thebibliography}

\label{lastpage}

\end{document}